\documentclass[12pt,superscriptaddress]{article}
\usepackage{palatino,cite,epsfig,amsmath,amssymb}

\begin{document}

\hyphenation{Higgs-strah-lung}


\renewcommand{\thefootnote}{\fnsymbol{footnote}}

\begin{titlepage}

\begin{flushright}
CERN-PH-TH/2004-058\\
Edinburgh 2004/08\\
IFT-2004-16\\
IISc-CHEP-2/04\\[0.5cm]
hep-ph/0404024\vspace{0.5cm}\\
\end{flushright}

\vspace{1cm}

\begin{center}
{\Large \bf CP Studies of the Higgs Sector:\\[0.3cm]
\large A contribution to the LHC / LC Study Group document\footnote{This report
is a slightly modified version of the contribution to the LHC / LC
Study Group document.}}\\[1cm]
{\large R.M.~Godbole$^{1}$, S.~Kraml$^{2,3}$, M.~Krawczyk$^{4}$, D.J.~Miller$^{5}$,\\
 P.~Nie\.zurawski$^{6}$ and A.F.~\.Zarnecki$^{6}$}\\[1cm]
{\it $^1$ Centre for Theoretical Studies, Indian Institute of Science, Bangalore 560012, India\\
     $^2$ Inst.\ f.\ Hochenergiephysik, \"Osterr.\ Akademie d.\ Wissenschaften, 1050 Wien, Austria\\
     $^3$ CERN, Physics Department, Theory Div., 1211 Geneva 23, Switzerland\\
     $^4$ Institute of Theoretical Physics, Warsaw University, 00-681 Warsaw, Poland\\
     $^5$ School of Physics, The University of Edinburgh, Edinburgh EH9 3JZ, Scotland\\
     $^6$ Institute of Experimental Physics, Warsaw University, 00-681 Warsaw, Poland}\\
\end{center}

\vspace{3cm}

\begin{abstract}
\noindent
The CP structure of the Higgs sector will be of great interest to
future colliders. The measurement of the CP properties of candidate
Higgs particles will be essential in order to distinguish models of
electroweak symmetry breaking, and to discover or place limits on
CP-violation in the Higgs sector.  In this report we briefly summarize
various methods of determining the CP properties of Higgs bosons at
different colliders and identify areas where more study is
required. We also provide an example of a synergy between the LHC, an
$e^+e^-$ Linear Collider and a Photon Collider, for the examination of
CP-violation in a Two-Higgs-Doublet-Model.
\end{abstract}

\end{titlepage}


\subsection*{1~~~~Introduction}

Discovery of the Higgs boson will be one of the primary goals of the
next generation of colliders. If, as hoped, one or more ``Higgs boson
like'' particles are observed, the next task will be to measure their
masses and quantum numbers and identify whether they are the Higgs
bosons of the Standard Model (SM), a Two-Higgs-Doublet-Model (2HDM),
the Minimal Supersymmetric Standard Model (MSSM), or some more exotic
alternative. In particular, the CP quantum numbers of the Higgs
boson(s) provide good discrimination, and consequently CP studies in
the Higgs sector will be a major focus when the physics programme of
the LHC \cite{atlas,cms}, an $e^+e^-$ Linear Collider
\cite{Aguilar-Saavedra:2001rg} or a Photon Collider
\cite{Badelek:2001xb} is mature.  Such studies of the CP properties of
the Higgs sector will involve establishing the CP eigenvalue(s) for
the Higgs state(s) if CP is conserved, and measuring the mixing
between the CP-even and CP-odd states if it is not. CP violation in
the Higgs sector~\cite{Weinberg:1976hu}, possible in multi-Higgs
models, is indeed an interesting option to generate CP violation
beyond that of the SM, possibly helping to explain the observed Baryon
Asymmetry of the Universe~\cite{Dine:2003ax}.
  
In order to identify the CP nature of a Higgs boson, one must probe
the structure of its couplings to known particles, in either its
production or decay. At tree level, the couplings of a neutral Higgs
boson $\phi$, which may or may not be a CP eigenstate,\footnote{For CP
eigenstates, a pure scalar will be denoted by $H$ and a pure
pseudoscalar by $A$. Otherwise we use the generic notation $\phi$.}
to fermions and vector bosons can be written as
\begin{equation}
   f\bar f\phi:~-\bar f(v_f+ia_f\gamma_5)f\,\frac{gm_f}{2m_W},\qquad
   VV\phi:~c_V\,\frac{gm_V^2}{m_W}\,g_{\mu\nu}\,
\label{eq:sec24-1}
\end{equation}
where $g$ is the usual electroweak coupling constant; $v_f$, $a_f$
give the Yukawa coupling strength relative to that of a SM Higgs
boson, and $c_V$ ($V=W,\,Z$) are the corresponding relative couplings
to gauge bosons\footnote{In principle, the $VV\phi$ coupling could
also contain an additional pseudoscalar coupling, although this is
absent in the SM and 2HDMs at tree-level (see later).}.  In the SM,
for a CP-even Higgs $v_f=c_V=1$ and $a_f=0$. A purely CP-odd Higgs
boson has $v_f=c_V=0$ and $a_f\not=0$, with the magnitude of $a_f$
depending on the model.  In CP-violating models, $v_f$, $a_f$ and
$c_V$ may all be non-zero at tree level. In particular, in the case of
a general 2HDM or the MSSM with CP violation, there are three neutral
Higgs bosons $\phi_i$, $i=1,2,3$, which mix with each other and share
out between them the couplings to the $Z$, $W$ and fermions; various
sum rules are given in
\cite{Gunion:1990kf,Gunion:1997aq,Ginzburg:2002wt}.  Due to this fact,
limits on the MSSM (and 2HDM) Higgs sector implied by LEP data are
strongly affected by the presence of CP violation
\cite{Gunion:1997aq,Carena:2002bb,data2003}.

In most formulations of CP-violating Higgs sectors~\cite{Dedes:1999sj,
Pilaftsis:1999qt,Choi:2000wz,Carena:2002bb,Ginzburg:2002wt,Dubinin:2002nx} 
the amount of CP mixing is small, being generated at the loop level, 
with only one of the couplings to gauge bosons or fermions sizable. 
In most cases, the predicted CP mixing is also a function of the CP-conserving
parameters of the model, along with the CP-violating phases.%
\footnote{For the MSSM with explicit CP violation, computational tools 
for the Higgs sector are available\cite{Heinemeyer:2001qd,Lee:2003nt}.}
Thus observation and measurement of this mixing at the LC 
may give predictions for LHC physics; for instance for sparticle 
phenomenology in the MSSM.
Moreover, experiments at different colliders have different sensitivities 
to the various couplings of eq.~\ref{eq:sec24-1}. Hence a combination of
LHC, LC and photon collider (PLC) measurements of both CP-even and CP-odd 
variables may be necessary to completely determine the coupling structure 
of the Higgs sector. 
These are two ways in which the high potential of LHC-LC synergy 
for CP studies can be realised.

In what follows, we give an overview of the LHC, LC, and PLC potentials 
for CP studies in the Higgs sector. An example of  the LHC-LC synergy is 
presented as well.

\subsection*{2~~~~CP Studies at the LHC}

There are several ways to study the CP nature of a Higgs boson at the LHC. 
In the resonant s-channel process $gg \to\phi\to f\bar f$, the scalar or 
pseudoscalar nature of the Yukawa coupling gives rise to $f\bar f$ spin-spin 
correlations in the production plane \cite{Bernreuther:1997gs}.
A more recent study \cite{Khater:2003wq} looks at
this process in the context of a general 2HDM.

In the process $gg\to t\bar t\phi$, the large top-quark mass enhances the 
$v^2-a^2$ contribution, allowing a determination of the CP-odd and CP-even 
components of a light Higgs Boson~\cite{Gunion:1996xu,Field:2002gt}. 
While this method should provide a good test for verifying a pure scalar 
or pseudoscalar, examination of a mixed CP state would be far more 
challenging, requiring $600~{\rm fb}^{-1}$ to distinguish an equal 
CP-even/CP-odd mixture at $\sim 1.5\,\sigma$~\cite{Gunion:1996xu}.

Higgs decay into two real bosons, $\phi\to ZZ$, with $Z\to l^+l^-$,
~\cite{Choi:2002jk,Buszello:2002uu} can be used to rule out
a pseudoscalar state by examining the azimuthal or polar angle distributions 
between the decay lepton pairs. Below the threshold, $\phi \to Z^*Z$, 
extra information is provided by the threshold behaviour of the virtual 
$Z$ boson invariant mass spectrum. 
This way,  one could rule out a pure $0^-$ state at $>5\sigma$
with $100~{\rm fb}^{-1}$ in the SM.  
An extension of these studies to scalar-pseudoscalar mixing is under progress.

In weak boson fusion, the Higgs boson is produced in association with two 
tagging jets, $qq \to W^+W^-qq \to \phi qq$. As with the decay to $ZZ$, 
the scalar and pseudoscalar couplings lead to very different azimuthal 
distributions between the two tagging jets~\cite{Plehn:2001nj}.
A similar idea may be employed in $\phi + 2 jets$ 
production~\cite{DelDuca:2001ad} in gluon fusion.  Higher order 
corrections~\cite{Odagiri:2002nd} may, however, reduce this correlation 
effect strongly.

Another approach uses the exclusive (inclusive) double diffractive process 
$pp\to p+\phi+p$ ($pp\to X+\phi+Y$)
\cite{Khoze:2001xm,Cox:2003xp,Khoze:2004rc} 
with large rapidity gaps between the $\phi$ and the (dissociated) protons. 
The azimuthal angular distribution between the tagged forward protons 
or the transverse energy flows in the fragmentation regions reflect 
the CP of the $\phi$ and can be used to probe CP mixing. 
This process is particularly promising for the region $m_\phi<60$~GeV, 
in which a Higgs signal may have been missed at LEP due to CP violation.

\subsection*{3~~~~CP Studies at an \boldmath $e^+e^-$ Linear Collider}

In $e^+e^-$ collisions, the main production mechanisms of neutral
Higgs bosons $\phi$ are 
(a) Higgsstrahlung $e^+e^-\to Z\phi$, 
(b) $WW$ fusion $e^+e^-\to \phi\,\nu\bar\nu$, 
(c) pair production $e^+e^-\to \phi_i\,\phi_j$ ($i \neq j$) and 
(d) associated production with heavy fermions, $e^+e^-\to f\bar f\phi$.  
Studies of CP at the Linear Collider aim at extracting the relevant
couplings mentioned in eq.~\ref{eq:sec24-1}.  Recall that a pure
pseudoscalar of the 2HDM or MSSM does not couple to vector bosons at tree
level.  The observation of all three $\phi_i$ $(i=1,2,3)$ in a given
process, e.g. $e^+e^-\to Z\phi_{1,2,3}$, therefore represents evidence
of CP violation
\cite{Mendez:1991gp,Grzadkowski:1999ye,Akeroyd:2001kt}.

In the Higgsstrahlung process, if $\phi$ is a pure scalar  the $Z$
boson is produced in a state of longitudinal polarization at high
energies \cite{Barger:1993wt,Hagiwara:1993sw}. For a pure pseudoscalar, 
the process proceeds via loops and the $Z$ boson in the final state is 
transversally polarized. The angular distribution of $e^+e^-\to ZH$ is 
thus $\propto\sin^2\theta_Z$, where $\theta_Z$ is the production angle 
of the $Z$ boson w.r.t.\ to the beam axis in the lab frame, while that 
of $e^+e^-\to ZA$ is $\propto(1+\cos^2\theta_Z)$. A forward-backward
asymmetry would be a clear signal of CP violation.  
Furthermore, angular correlations of the $Z\to f\bar f$ decay can be used 
to test the $J^{PC}$ quantum numbers of the Higgs boson(s). Measurements of
the threshold excitation curve can give useful additional information
\cite{Miller:2001bi,Dova:2003py}. A study in
\cite{Aguilar-Saavedra:2001rg} parametrised the  effect of CP violation
by adding a small $ZZA$ coupling with strength $\eta$ to 
the SM matrix element,  ${\cal M } = {\cal M}_{ZH} + i \eta  {\cal M}_{ZA}$,~
and showed that $\eta$ can be measured to an accuracy of $3.2\%$ with 
$500$ fb$^{-1}$.

Angular correlations of Higgs decays can also be used to determine the
CP nature of the Higgs boson(s), independent of the production
process; see 
\cite{Kramer:1993jn,Grzadkowski:1995rx,Gunion:1996vv} 
and references therein. 
The most promising channels are $\phi\to\tau^+\tau^-$ ($m_\phi<2m_W$) and 
$\phi\to t\bar t$ ($m_\phi>2m_t$) which in contrast to decays into $WW$ or 
$ZZ$ allow equal sensitivity to the CP-even and CP-odd components of $\phi$.  
  
A detailed simulation of $e^+e^-\to ZH$ followed by $H\to\tau^+\tau^-$ and 
$\tau^\pm\to\rho^\pm\bar\nu_\tau(\nu_\tau)$ 
\cite{Bower:2002zx,Desch:2003mw,Worek:2003zp} 
showed that CP of a 120~GeV SM-like Higgs boson can be measured to 
$\ge 95\%$ C.L. at a 500~GeV $e^+e^-$ LC with 500~fb$^{-1}$ of luminosity.
In case of CP violation, the mixing angle between the scalar and pseudoscalar 
states may be determined to about 6 degrees~\cite{Desch:2003rw}, the
limiting factor being statistics.

\subsection*{4~~~~CP Studies at a Photon Collider}

A unique feature of a PLC is that two photons can form a $J_z = 0$ state 
with both even and odd CP. As a result a PLC has a similar level of 
sensitivity for both the CP-odd and CP-even components of a CP-mixed state:
\begin{equation}
  {\rm CP\!-\!even:}
  \epsilon_1\cdot \epsilon_2 = -(1+\lambda_1\lambda_2)/2 , \quad
  {\rm CP\!-\!odd:}
  [\epsilon_1 \times \epsilon_2] \cdot k_{\gamma}
  =\omega_{\gamma} i \lambda_1(1+\lambda_1\lambda_2)/2,
\end{equation}
$\omega_i$ and $\lambda_i$ denoting the energies and  helicities of the 
two photons respectively; the helicity of the system is equal to 
$\lambda_1-\lambda_2$.
This contrasts the $e^+e^-$ case, where it is easy to discriminate between 
CP-even and CP-odd particles but may be difficult to detect small CP-violation 
effects for a dominantly CP-even Higgs boson~\cite{Hagiwara:2000bt}.  
For the PLC, one can form three polarization asymmetries in terms of helicity 
amplitudes which give a clear measure of CP mixing \cite{Grzadkowski:1992sa}. 
In addition, one can use information on the decay products of $WW$, $ZZ$, 
$t\bar t$ or $b\bar b$ coming from the Higgs decay.
Furthermore, with circular beam polarization almost mass degenerate 
(CP-odd) $A$ and (CP-even) $H$ of the MSSM may be separated 
\cite{Muhlleitner:2001kw, Niezurawski:2003ir,Asakawa:1999gz}.

A measurement of the spin and parity of the Higgs boson may also be
performed using the angular distributions of the final-state fermions
from the Z boson decay, which encode the helicities of Z's.   A
detailed study was performed for above and below  the ZZ threshold in
\cite{Choi:2002jk}. A realistic simulation based on this analysis was
made recently in~\cite{Niezurawski:2003ik}.

The same interference effects as mentioned above can be used in the process 
$\gamma \gamma \rightarrow \phi \rightarrow t \bar t$ 
\cite{Asakawa:2000jy,Godbole:2002qu} 
to determine the $t \bar t\phi$ and $\gamma \gamma \phi$ couplings 
for a $\phi$ with indefinite CP parity.

\subsection*{5~~~~Example of LHC-LC synergy}

As an example of the LHC-LC synergy, we consider the SM-like, type II 2HDM
with CP-violation~\cite{Niezurawski:2003ik,2HDM}.
%
We study production of  $\phi_2$ in the mass range 200 to 350~GeV, 
decaying to $VV$, $V = W/Z$, at the LHC, LC and PLC. In particular,
we investigate the interplay of different experiments for the determination 
of $\tan \beta$ and the CP mixing angle $\Phi_{HA}$.

\begin{figure}[h]
     \epsfig{figure=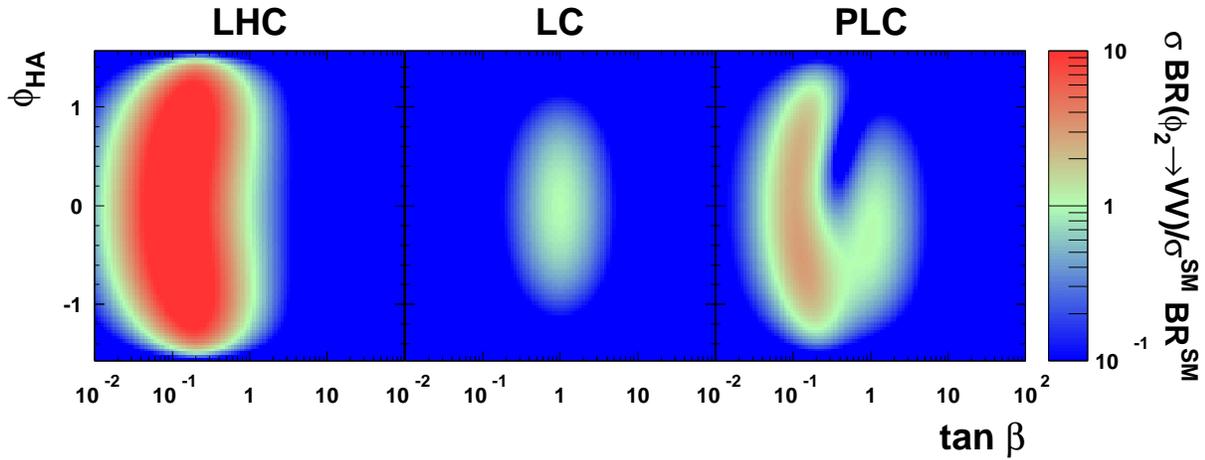,width=\textwidth,clip=}
\vspace{-1cm}
\caption{ $\sigma \times$ BR for $\phi_2 \rightarrow VV$ with $V =
W/Z$, relative to the SM expectation, for a mass of 250 GeV, as a
function of $\tan \beta$ and $\Phi_{HA}$ for the LHC, LC and PLC.}
 \label{nzk:cros2}
 \end{figure}

Figure~\ref{nzk:cros2} shows the expected rates for $\phi_2$ with 
$m_{\phi_2}=250$ GeV relative to the SM ones, as a function of 
$\tan\beta$ and  $\Phi_{HA}$.  For a SM Higgs boson, 
the expected precision on $\sigma \times BR(H \rightarrow VV)$ is 
$\sim 15$\% at the LHC \cite{cms_tn_95-018,Kinnunen:2002cr} and 
better than  10\%  at a LC and PLC \cite{lc-phsm-2003-066,nzk_wwzz}. 
A PLC will  allow to measure $\Gamma_{\gamma \gamma}$ with a precision 
of 3--8\%  and the phase of the $\phi \rightarrow \gamma \gamma$ 
amplitude, $\Phi_{ \gamma \gamma}$, to $40-120$~mrad \cite{nzk_wwzz}.   

\begin{figure}[h]
  \begin{center}
     \epsfig{figure=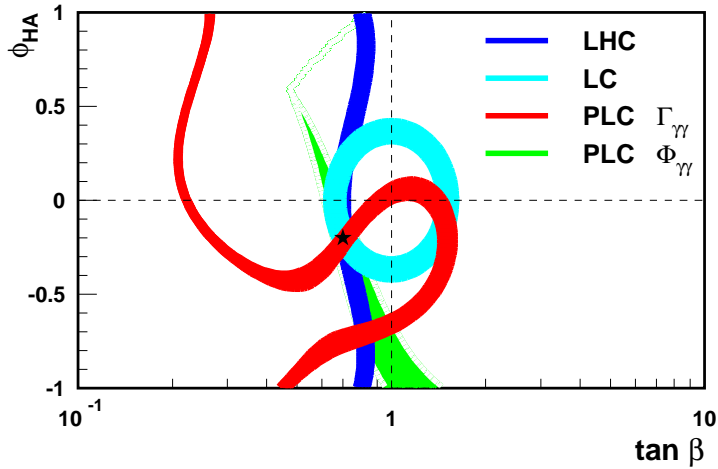,width=10cm,clip=}
  \end{center}
\vspace{-1cm}
\caption{ 
1-$\sigma$ bands for the determination of  $\tan \beta$ and  $\Phi_{HA}$ 
from  measurements at the LHC, LC and PLC, for the case $\tan \beta = 0.7$ 
and $\Phi_{HA} = -0.2$. The assumed parameter values are  indicated by 
a star ($\star$).}
 \label{nzk:syn2} 
 \end{figure}

Figure~\ref{nzk:syn2} shows the $1\,\sigma$ bands for determination of
$\tan \beta$ and $\Phi_{HA}$, at the LHC, LC and PLC for a particular
choice of parameters : $\tan \beta=0.7$ and $\Phi_{HA}=-0.2$. The
chosen point is indicated by a star. For the PLC, information from
$\Gamma_{\gamma\gamma}$ and $\Phi_{\gamma \gamma}$ is included. As can
be seen, an accurate determination of both parameters of the model
requires a combination of data from all three colliders.
%

\subsection*{6~~~~Summary}

The LHC, an $e^+e^-$ LC, and a LC in the photon collider option (PLC)
will be able to provide important information on the CP quantum
numbers of the Higgs boson(s).  We have summarised the potentials of
the different colliders in this document and discussed the possible
LHC-LC synergy.

In the MSSM, for instance, the size of CP-violating effects in the Higgs
sector depends in part on the sparticle spectrum.  
Observation and measurement of Higgs-sector CP mixing at the LC 
can hence give predictions for phenomenology at the LHC in the 
CP-conserving sector, thus providing a high potential of LHC-LC synergy.  
A detailed study of this issue is, however, still missing.

Moreover, experiments at different colliders have different sensitivities
to the various couplings of eq.~\ref{eq:sec24-1}. Hence a combination of
LHC and LC/PLC measurements of both CP-even and CP-odd variables may be
necessary to completely determine the coupling structure of the Higgs sector.
In this document we have presented a first analysis which exemplifies this
realisation of LHC-LC synergy. While the example presented shows a high 
potential of the LHC-LC synergy for CP studies, detailed realistic simulations
still need to be performed.

\subsection*{Acknowledgments}

This work was partially supported by the Polish Committee for
Scientific Research, grant no.~1~P03B~040~26 and project
no.~115/E-343/SPB/DESY/P-03/DWM517/2003-2005. \linebreak[4]
R.G. would like to acknowledge the support of Department of Science
and Technology, India, under project number SP/S2/K-01/2000-II.



\end{document}